\documentclass[referee,useAMS,usenatbib]{mn2e}
\usepackage{graphicx}
\usepackage{hyperref}
\usepackage{epstopdf}
\usepackage{bm}
\usepackage{lscape}

\title[PARAS SPEC]
{Spectral Analysis Code: PARAS SPEC}

\author[Chaturvedi et al.]
  {Priyanka Chaturvedi,
    Abhijit Chakraborty
   and B.G. Anandarao
\\
 Astronomy \& Astrophysics Division, Physical Research Laboratory, Ahmedabad 380009, India\\ }
\date{Received 2016 XXXX XXX }

\pagerange{\pageref{firstpage}--\pageref{lastpage}} \pubyear{2016}

\def\LaTeX{L\kern-.36em\raise.3ex\hbox{a}\kern-.15em
    T\kern-.1667em\lower.7ex\hbox{E}\kern-.125emX}

\newcommand{\aap}{    {\it Astron. Astrophys.}}
\newcommand{\aaps}{   {\it Astron. Astrophys. Suppl.}}
\newcommand{\aapr}{   {\it Astron. Astrophys. Rev.}}

\newcommand{\aj}{     {\it Astron. J.}}
\newcommand{\apj}{    {\it Astrophys. J.}}

\newcommand{\apjs}{   {\it Astrophys. J. Supplement.}}

\newcommand{\mnras}{  {\it Mon. Not. Roy. Astron. Soc.}}

\newcommand{\pasp}{   {\it Pub. Astron. Soc. Pac.}}

\newcommand{\ssr}{    {\it Space Sci. Rev.}}

\newcommand{\araa}{{\it ARA\&A}}

\newcommand{\gca}{{\it Geochimica et Cosmochimica Acta}}

\begin{document}
\label{firstpage}
\maketitle

\begin{abstract}
The light emitted from the stellar photosphere serves as a unique signature for the nature of stars. The behaviour of these stellar lines depend upon the surface temperature, mass, evolutionary status and chemical composition of the star. With the advent of high-resolution spectrographs coupled with medium to large aperture telescopes around the globe, there is plenty of high-resolution and high signal-to-noise ratio data available to the astronomy community. Apart from radial velocity (RV) studies, such data offer us the unique opportunity to study chemical composition and atmospheric properties of the star. The procedure used to derive these parameters must be automated and well adaptable to data available from any high-resolution spectrograph. We hereby present an IDL code, \textit{PARAS SPEC}, which was primary designed to handle high-resolution spectroscopy data from PARAS spectrograph coupled with the 1.2~m telescope at Mt. Abu, India. This code is designed to adapt with data from other spectrographs as well. The code \textit{PARAS SPEC} estimates the stellar atmospheric parameters from the analysis of stellar spectra based on two primary methods, synthetic spectral fitting and equivalent width method. Synthetic spectral fitting method involves fitting of the observed spectrum with different synthetic spectra for a set of stellar parameters. The second method is based on equivalent widths (EWs) that are used to derive abundances for a set of Fe~I and Fe~II lines from the observed spectra. The detailed methodology used to design this code and comparison of the results from literature values are presented in this paper.
 
 \end{abstract}
 
 \begin{keywords}
 	stars : atmospheres -- methods: data analysis --	 
 	techniques: spectroscopic
 \end{keywords}

\section{Introduction}

Double-lined EBs (SB$_2$) enable the most accurate measurements of the radii and masses of the stars. Since in such systems, spectra of both the stars can be recorded, radial velocity (RV) measurements lead to precise determination of masses of the stars in the system directly. However, in single-lined EBs (SB$_1$) and star-planet systems, where the spectra of only the primary source is available, mass of the secondary or the planet is deduced by quantifying the amplitude of the wobble on the primary star in the binary system. A detailed study by \cite{Torres2010} on a large sample of EBs has led to the establishment of an empirical relation for the mass and radius of the stars above 0.6~M$_{\odot}$ based on the stellar parameters, i.e, ${\rm log}~g$, the $T_{\rm{eff}}$ and $[Fe/H]$. There are many host stars for which spectral properties have not been studied so far and therefore it is of utmost importance to derive the same in order to draw inference on their mass and radius. 

In this context, we have developed a pipeline, \textit{PARAS SPEC}, to estimate the stellar atmospheric parameters from the analysis of stellar spectra. With high-resolution spectroscopy, it is possible to determine $T_{\rm{eff}}$ and $[Fe/H]$ (based on Fe~I and Fe~II lines) as well as ${\rm log}~g$ (based on Mg~I lines) at high accuracies. The pipeline, a set of {\tt{IDL}}-based tools, is developed to facilitate the determination of stellar properties. Though it was primarily designed to facilitate stellar property estimation from PARAS data, it is easily possible to adapt to spectrographs of different resolutions. The high-resolution spectroscopy data presented here are from PARAS and SOPHIE spectrographs. Some of the observed stars are F, G and K type stars wich are part of the \textit{Gaia} ESO survey \citep{Gilmore2012} and others are part of the PARAS exoplanet and EB program \citep{Chakraborty2014, Chaturvedi2016}.  The technique is based on two principal methods. The first method involves fitting of the observed spectrum with different synthetic spectra for a set of stellar parameters. The best-match having minimum $\chi^2$ value between the observed and the synthetic spectra gives the best-fit values for the stellar parameters of the star. The second method is based on equivalent widths (EWs) that are used to derive abundances for a set of Fe~I and Fe~II lines from the observed spectra \citep{Blanco-Cuaresma2014}. The abundances determined must fulfill the conditions of excitation equilibrium and ionization balance. The stellar model for which the conditions of equilibria are satisfied is considered to be the best-fit model for representing the observed spectra. 

We describe the \textit{PARAS SPEC} code in this paper by discussing both the methods individually. A comparison has been made by the results obtained from this code with those from literature based values.  The final section lists the conclusion and future upgrades of this work.

\section{Methodology}

The {\tt{IDL}}-based tool, named \textit{PARAS SPEC} is designed to facilitate the estimation of stellar parameters from high-resolution spectra. The \textit{PARAS SPEC} code requires blaze-corrected and normalized stellar spectra as an input. 
%The blaze-corrected and normalized stellar spectrum is an output of the data reduction and analysis pipeline, PARAS PIPELINE as described in Ch~\ref{ch2}. 
The detailed steps used by \textit{PARAS SPEC} are described as follows: 

\subsection{High-resolution spectroscopy data} \label{blaze}
\textit{PARAS SPEC} has been designed with the primary motivation to determine atmospheric properties of the stars from the data obtained from Physical Research Laboratory Advanced Radial velocity Abu sky Search (PARAS) spectrograph. PARAS is an optical fiber-fed high-resolution (R~$\sim$~67000) cross-dispersed echelle spectrograph commissioned at the Mount Abu 1.2~m telescope in India (latitude: 24$^{\circ}$ 39$^{'}$ 10$^{''}$N, longitude: 72$^{\circ}$ 46$^{'}$ 47$^{''}$E, altitude 1680~m). The spectrograph has a spectral coverage of $3800-9000~\AA$. However, for precise RV measurements, wavelength range of $3800-6800~\AA$~is utilized with Thorium Argon (ThAr) simultaneous calibration method \citep{Chakraborty2014}. \textit{PARAS SPEC} is designed in such a way that it is suitable to adapt to data from other spectrographs as well. We have demonstarted this by presenting results obtained from using SOPHIE data in the following section. The echelle spectra obtained from PARAS are blazed at each order, which needs to be accounted for before the spectra is normalized. For this purpose, a polynomial function is fitted iteratively to an accuracy of $\sim$1~$\%$ across the stellar continuum blaze profile for each order after ignoring absorption features in the stellar spectra. The observed spectra are then divided by this function to blaze correct and normalize it at a given epoch. For SOPHIE data, we retrieved the normalized reduced data from the archive.

\subsection{A library of synthetic spectra}

Observed high-resolution spectra are compared with this reference library spectra which is at a similar resolving power as that of the observed spectra. A library of synthetic spectra is generated using the code {\tt{SPECTRUM}} \citep{Gray1999}~\footnote{http://www.appstate.edu/$\sim$grayro/spectrum/spectrum276/spectrum276.html}. Initially, PARAS observed solar spectrum was used to fix parameters, such as micro-turbulent velocity ($v_{\rm{micro}}$), macro-turbulent velocity ($v_{\rm{macro}}$) and stellar abundances for the synthetic spectral library. The relevant details are briefly described as follows.

\subsubsection{{\tt{SPECTRUM}} program} Synthetic spectra generator code {\tt{SPECTRUM}} utilizes the Kurucz models \citep{Kurucz1993} for stellar atmosphere parameters. {\tt{SPECTRUM}} works on the principle of local thermodynamic equilibrium  and plane parallel atmospheres. It is suitable for generation of synthetic spectra for stars from B to mid M type. It is developed with C compiler environment with a terminal mode interface to access it. 

\subsubsection{Solar spectrum} A solar spectrum (SNR~$\sim$150) observed with PARAS is taken up as a first step to fix various parameters for the library of synthetic spectra. The synthetic spectra can be generated at a fine wavelength spacing of 0.01~\AA~between two consecutive wavelength values. This resolving power (R~$\sim$~500,000 at 5000~\AA) is very high in comparison to the resolving power of PARAS (R~$\sim$~68,000 at 5000~\AA). For PARAS, the wavelength dispersion is 0.02~$\AA/$pixel at $5500~\AA$ and 0.024~$\AA/$pixel at $6500~\AA$. PARAS has a resolution element of 4~pixels and thus the FWHM of the spectral profile of PARAS at the central wavelength region, 5500~\AA, is $\sim$~0.08~\AA. In order to match the spectra, both the synthetic and observed spectra must have the same sampling and should be at the same resolving power. FWHM varies across the entire wavelength region covered by the PARAS spectra but for simplicity we use the the central wavelength FWHM for convolution of synthetic spectra with a Gaussian function. This is kept fixed for the entire library of synthetic spectra. All the parameters for the observed solar spectrum, such as, $v_{\rm{micro}}$, $v_{\rm{macro}}$, rotational velocity ($v \sin i$), $T_{\rm{eff}}$, $[Fe/H]$ and ${\rm log}~g$ are kept free. When all the parameters are kept free, the best-derived model having the least $\chi^2$ for the PARAS observed solar spectra has the following values for various parameters: $v_{\rm{micro}}$~=~0.85~km~s$^{-1}$ and $v_{\rm{macro}}$~=~2~km~s$^{-1}$.
%, $T_{\rm{eff}}$~=~5800 K, $[Fe/H]$~=~0.0, ${\rm log}~g$~=~4.1, $v~\sin i$~=3~kms$^{-1}$. 
The value for $v_{\rm{micro}}$ obtained here is in close agreement with the one derived by \cite{Blackwell1984}. The value of $v_{\rm{macro}}$~=~2~km~s$^{-1}$ derived here is consistent with the value of 2.18~km~s$^{-1}$ obtained by \citep{Valenti1996}. The value of $v_{\rm{micro}}$ and $[Fe/H]$ are partially degenerate as studied by \cite{Valenti2005}, which means there can be more than one combination of $v_{\rm{micro}}$ and $[Fe/H]$ for a best-fit solution. Thus, following a similar approach as that of \cite{Valenti2005}, it was decided to keep $v_{\rm{micro}}$ fixed in order to minimize the errors on estimation of $[Fe/H]$. A similar degeneracy is seen in the parameters $v_{\rm{macro}}$ and  $v \sin i$. Stars having temperatures between 5000~--~6500 K, similar to those observed with PARAS, have a range of $v_{\rm{macro}}$ values between $2-5$~km~s$^{-1}$ \citep{Valenti2005}. We set $v_{\rm{macro}}$ fixed at 2~km~s$^{-1}$ so that an upper limit on $v \sin i$ is determined, similar to the approach followed by \citep{Valenti2005}. There are two atomic line-lists present in the distribution of {\tt{SPECTRUM}}, \textit{luke.lst} and \textit{luke.iso.lst}.
We used the second line list for the entire library since it was found that it has relatively smaller $\chi^2$ than the first linelist (luke.lst) for the same synthetic models.

\subsubsection{The synthetic library} The synthetic library consists of the following building blocks. The tabulated stellar library generated is given as Table~\ref{library}.

\begin{itemize}

\item{\textbf{Models:}} The models for the synthetic spectra are retrieved from the Kurucz model database \citep{Kurucz1993}. The models are a byproduct of the larger combined family of models, the supermodels. Each supermodel belongs to a single metallicity and comprises different models having varying temperatures and surface gravity. Each model has a typical format, which the {\tt{SPECTRUM}} routine undertakes into consideration while execution. It consists of the following columns.

\begin{enumerate}
\item $\int$$\rho$~dx: mass depth (g~cm$^{-2}$)
\item $T\rm{_{eff}}$: temperature (K)
\item $P_{\rm{gas}}$: gas pressure (dynes~cm$^{-2}$)
\item $n_e$: electron density (cm$^{-3}$)
\item $\kappa_R$: Rosseland mean absorption coefficient (cm$^{2}$~gm$^{-1}$)
\item $P_{\rm{rad}}$: radiation pressure (g~cm$^{-2}$)
\item $v_{\rm{micro}}$: microturbulent velocity (m~s$^{-1}$)	
\end{enumerate}

different combinations of $T_{\rm{eff}}$, $[Fe/H]$, ${\rm log}~g$ and $v \sin i$ lead to generation of 19,200 synthetic spectra. The ranges of these parameters chosen are given in Table~\ref{library}.

\item{\textbf{Interpolation of model atmospheres:}}

The {\tt{SPECTRUM}}-generated synthetic library originally consists of a coarse grid in $T_{\rm{eff}}$, $[Fe/H]$ and ${\rm log}~g$.  A finer grid is required to achieve the close resemblance of the observed spectra with the synthetic spectra. Thus, during the course of execution of the synthetic spectral fitting routine, the synthetic models are interpolated in the desired range of $T_{\rm{eff}}$, $[Fe/H]$ and ${\rm log}~g$ to sharpen the precision of the derived parameters. The interpolation on the models is executed by the {\tt{IDL}} subroutine {\tt{kmod}}. The interpolated models then have a finer interval in $T_{\rm{eff}}$ (50 K), $[Fe/H]$ (0.1 dex) and ${\rm log}~g$ (0.1 dex) as mentioned in Table~\ref{library}. 

\item{\textbf{Atomic line list:}}
{\tt{SPECTRUM}} code distribution includes two line lists, namely, \textit{luke.lst} and \textit{luke.iso.lst}. As mentioned earlier, we tested both the atomic line lists on the solar spectra and found the \textit{luke.iso.lst} yield better results in terms of a better minimized $\chi^2$.

\item{\textbf{Abundances:}}
The standard solar abundances, which are provided by \cite{Anders1989, Grevesse1998, Asplund2005, Grevesse2007, Asplund2009}, are used for  generation of synthetic spectra with \textit{PARAS SPEC}. The abundance value of any element for a star is computed by taking a ratio of the abundance of that element present in the star with respect to the solar abundance for the same element. 
\begin{equation}
[Fe/H]=log_{10}\left(\frac{[Fe/H]_{star}}{[Fe/H]_{sun}}\right)
\label{eq1}
\end{equation}

\end{itemize}
%------------------------------------------------------------------------------------------------
%  TABLE 1
%------------------------------------------------------------------------------------------------

\begin{table}[!ht]
\centering
\caption[Synthetic Spectral Library]{Synthetic Spectral Library} 
\label{library}
\begin{tabular}{lccc}\\
\multicolumn{1}{l}{Parameter} & \multicolumn{1}{c}{Range} & \multicolumn{1}{c}{Original Interval} & \multicolumn{1}{c}{Interpolated interval}\\ 
\hline
$T_{\rm{eff}}$ (K) & $4000-7000$ & $250$ & $50$\\
$[Fe/H]$ (dex) & $-2.5-0.5$ & $0.5$ & $0.1$ \\
${\rm log}~g$ (dex) & $1.0-5.0$ & $0.5$ & $0.1$ \\
Wavelength (\AA) & $5050-6560$ & $0.01$ & $0.01$\\
$v sin i $ (km s$^{-1}$)  & $1-40$ & $1$ & $1$\\
\hline
\end{tabular}
\end{table}

%------------------------------------------------------------------------------------------------

\subsection{Preliminary considerations}

The spectra across various orders are blaze-corrected as discussed in \S~\ref{blaze} and are stitched together to produce a continuous single spectrum. Many such epochs are co-added in velocity space to improve the signal-to-noise ratio (SNR) of the stellar spectra. In absence of any absorption line, the synthetic spectral continuum value is 1.0 since it is a normalized spectra. For the observed spectra, the continuum level fluctuates around 1.0, as the blaze correction is not accomplished accurately for the case of deep absorption lines in the spectra. The continuum value in the vicinity of a prominent absorption feature is likely to be under-estimated due to the broad wings in the absorption profile. To eliminate this issue, we devised a method wherein the observed stellar spectra is normalized with respect to its stellar continuum in small wavelength intervals of $5$~\AA~each.

\subsection{Synthetic Spectral Fitting method}

%------------------------------------------------------------------------------------------------
%  FIG 1
%-----------------------------------------------------------------------------------------------

\begin{figure}[!ht]
\centering
\hbox{
\includegraphics[width=5.0in]{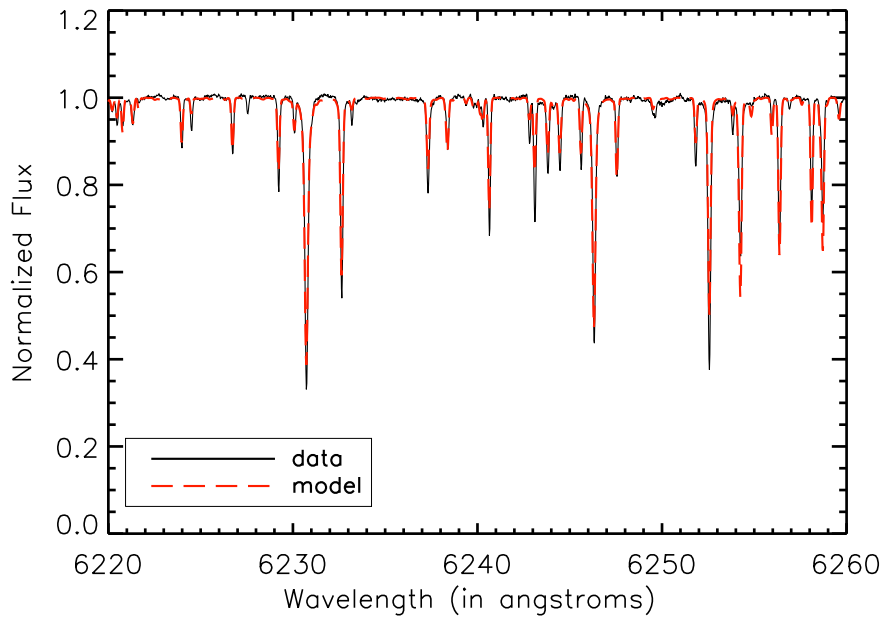}
}
\caption[Best-fit synthetic model from \textit{PARAS SPEC} overlaid on observed spectra for $\tau$~Ceti] {Observed normalized spectra for $\tau$~Ceti (solid black line) plotted across the wavelength region of $6220-6260~\AA$. Overplotted is the modelled spectra (red dash line) obtained from \textit{PARAS SPEC} analysis, with $T\rm{_{eff}}$ of 5400 K, $[Fe/H]$ of $-0.5$ and ${\rm log}~g$ of 4.4.}
\label{tauceti_spec}
\end{figure}

%----------------------------------

%------------------------------------------------------------------------------------------------
%  FIG 2
%-----------------------------------------------------------------------------------------------

\begin{figure}[!ht]
\centering
\hbox{
\includegraphics[width=5.0in]{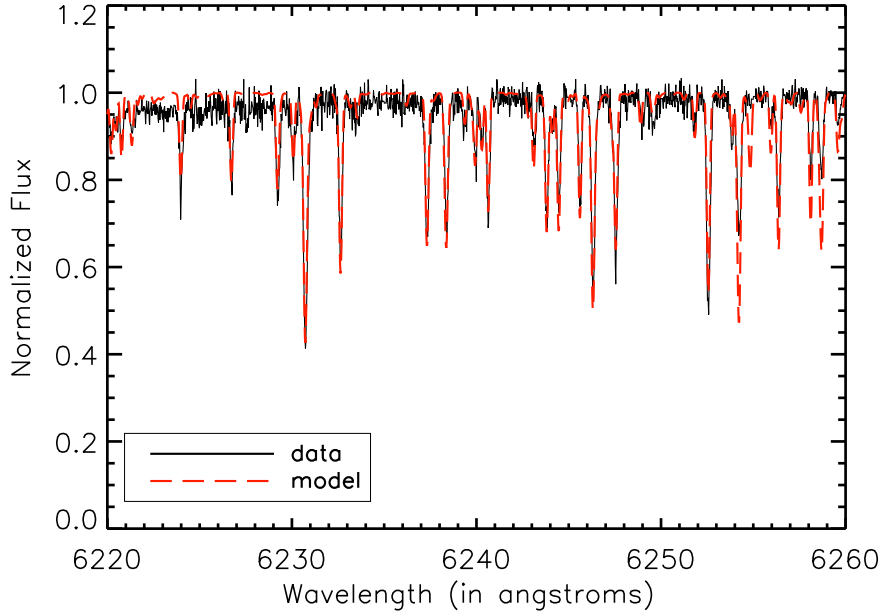}
}
\caption[Best-fit synthetic model from \textit{PARAS SPEC} overlaid on observed spectra for KID~5108214]{Observed normalized spectra for KID~5108214 (solid black line) plotted across the wavelength region of $6220-6260~\AA$. Overplotted is the modelled spectra (red dash line) obtained from \textit{PARAS SPEC} analysis, with $T\rm{_{eff}}$ of 6000 K, $[Fe/H]$ of $0.3$ and ${\rm log}~g$ of 4.0.}
\label{kid_spec}
\end{figure}

%-----------------------------------

The synthetic spectral fitting method is then executed in four steps. $T_{\rm{eff}}$, $[Fe/H]$ and ${\rm log}~g$ are all kept free in the first step.
The RMS residuals defined as $\sum\nolimits {(O(i)-M(i))}^2$ are computed between the observed and the synthetic spectra at each wavelength $\lambda _i$ in the wavelength range $5050-6000~\AA$. Here, O~$\&$~M are the observed and model spectra respectively. The best-match values of $T_{\rm{eff}}$ and $[Fe/H]$ are determined from this step. The first step of synthetic spectral fitting is executed on the set of models which have not been interpolated to a finer grid. The $T_{\rm{eff}}$, $[Fe/H]$ and ${\rm log}~g$ parameters determined from the original set of library (case of non-interpolated models) will thus have a coarse precision, as given in second column of Table~\ref{library}. 

In the second step, the parameters obtained from the previous step are used as an initial guess value and interpolation is done simultaneously at finer precision in $T_{\rm{eff}}$, $[Fe/H]$ and ${\rm log}~g$ in the wavelength range between $5050-6000~\AA$ as given in third column of Table~\ref{library}. The interpolation is done in the vicinity of the guess values on the three parameters obtained from the first step, i.e., $T_{\rm{eff}}$ in a range of $\pm$~250 K, $[Fe/H]$ in a range of $\pm$ 0.3 and ${\rm log}~g$ in a range of $\pm$ 0.3. The best-determined values derived from this second step are considered as initial approximations on stellar parameters for the third step. 

In the third step, the same routine is re-run, but this time only on the ${\rm log}~g$ sensitive Mg~I lines in the wavelength region of $5160-5190~\AA$ by keeping $T\rm{_{eff}}$ and $[Fe/H]$ the same as obtained from the second step. 

The fourth step is executed again on the wavelength region of $5160-5190~\AA$ on the interpolated models which are generated during the course of execution of the code. The best-match model determined at this step gives us the value for ${\rm log}~g$ along with previously determined values of $T_{\rm{eff}}$ and $[Fe/H]$ from the second step. 

A typical best-determined synthetic spectra from the \textit{PARAS SPEC} routine overlaid on normalized observed spectra for star $\tau$~Ceti is shown in Fig.~\ref{tauceti_spec}. The SNR for this spectra is $\sim$~500/pixel at 6000~$\AA$. The entire wavelength region of $5050-6500~\AA$~is covered for this star. In Fig.~\ref{kid_spec}, the spectra for a faint star (V=8 mag), KID~5108214, having a SNR of $\sim$~100~pixel$^{-1}$ at 6000~$\AA$ is shown. The observed spectra appears more noisy than $\tau$~Ceti due to relatively less SNR in comparison to that for $\tau$~Ceti. If the spectra having SNR below $\sim80$ are used, the uncertainties on each of the stellar parameters are approximately equal to the coarse grid size of the library (column 3 of Table~\ref{library}). Thus, for this method, we mention a lower limit on SNR of that being $\sim80$ in the entire wavelength region covered. For fainter stars having magnitudes between $7-11$ in V band, the method fails if applied as it is described above for the entire wavelength range $5050-6000~\AA$. Despite co-adding spectra for several epochs, SNR for some stars is below~80 in the blue end~($5050-6000~\AA$). Since, the CCD is more sensitive to redder wavelengths, even for stars having magnitudes between $7-11$ in V band, we expect a higher SNR (generally above 80) in the red portion of the spectra. Thus, we concentrate on the wavelength region between $6000-6500~\AA $ of the CCD where SNR is above~80 for such cases. However, despite the fact that we are able to determine  $T_{\rm{eff}}$ and $[Fe/H]$, we lose crucial information of the surface gravity, which is dependent on the Mg~I lines occurring between $5160-5190~\AA$. Thus, for such cases, it is necessary to rely upon the EW method inspired by the work of \cite{Blanco-Cuaresma2014}.

\subsection{EW Method}

The EW method works on the principle in which one seeks the neutral and ionized iron lines to satisfy the two equilibria, namely, excitation equilibrium and ionization balance. A set of neutral and singly ionized lines is acquired from the iron line list by \cite{Sousa2014}. The method is executed step by step as described below. 

\subsubsection{Measurement of EW}

The EW of a spectral line is dependent on the number of photons that are absorbed at a particular wavelength.

The EW defined as:
\begin{equation}
W_{\lambda}=\int(1-F_{\lambda}/F_o)d\lambda.
\label{eq2}
\end{equation}
Here, $F_o$ represents the continuum level, $F_\lambda$ represents flux at a given wavelength, $\lambda$ and $W_{\lambda}$ represents the EW at that $\lambda$. It can be geometrically represented as the area of the line profile. Hence, it can be represented as the width in wavelength of a rectangular profiled line 100$\%$ deep having the same area in a flux vs wavelength plot as the actual spectral line profile \citep{Emerson1996}. For the measurement of EW, a Gaussian function is fitted for each spectral line of the observed spectra corresponding to all the iron lines that are present in the line list given by \cite{Sousa2014}. Each Gaussian-fitted profile corresponding to the iron line list is then carefully inspected and poor fits and line blends are eliminated.

\subsubsection{Determination of abundance}

The {\tt{SPECTRUM}} code facilitates estimation of abundance of elements from their spectral lines. After careful inspection of the EW fits as discussed above, a set of EW of the fitted lines is given as an input to the {\tt{ABUNDANCE}} subroutine of {\tt{SPECTRUM}}. The subroutine uses various stellar models which are formed as a combination of different $T\rm{_{eff}}$, $[Fe/H]$, ${\rm log}~g$ and $v_{\rm{micro}}$. The output of the {\tt{ABUNDANCE}} routine is as follows.
\begin{center} 
6127.895~~26.0~~1.08~~-4.581~~7.459~~-0.041 
\end{center}
The first and second columns stand for the central wavelength of the line formed by an element and the atomic number of that element, the third column is the $v_{\rm{micro}}$ (in km~s$^{-1}$). The fourth column gives the abundance on the scale where total abundances are expressed with respect to the total number density of atoms (and ions), the fifth column expresses abundances on the normal scale, in which the logarithmic abundance of hydrogen is equal to 12.0. The last column is the abundance relative to the unscaled abundances with respect to solar abundances for the current case. 

\subsubsection{The three golden rules}
The main purpose of calculating EW and thereby abundances is the fact that the abundances of a given species follow a set of three golden rules. This fact can be employed to choose a best-fit model of synthetic spectra in which all the rules are simultaneously satisfied. These three rules are:
\begin{enumerate}
\item Abundances as a function of excitation potential (EP) should have no trends. 
\item Abundances as a function of reduced EW (EW/$\lambda)$ should exhibit no trends.
\item Abundances of neutral iron (Fe I) and ionized iron (Fe II) should be balanced. 
\end{enumerate}

%------------------------------------------------------------------------------------------------
%  FIG 4
%------------------------------------------------------------------------------------------------

\begin{figure}[!ht]
\centering
%\hbox{
\includegraphics[width=5.0in]{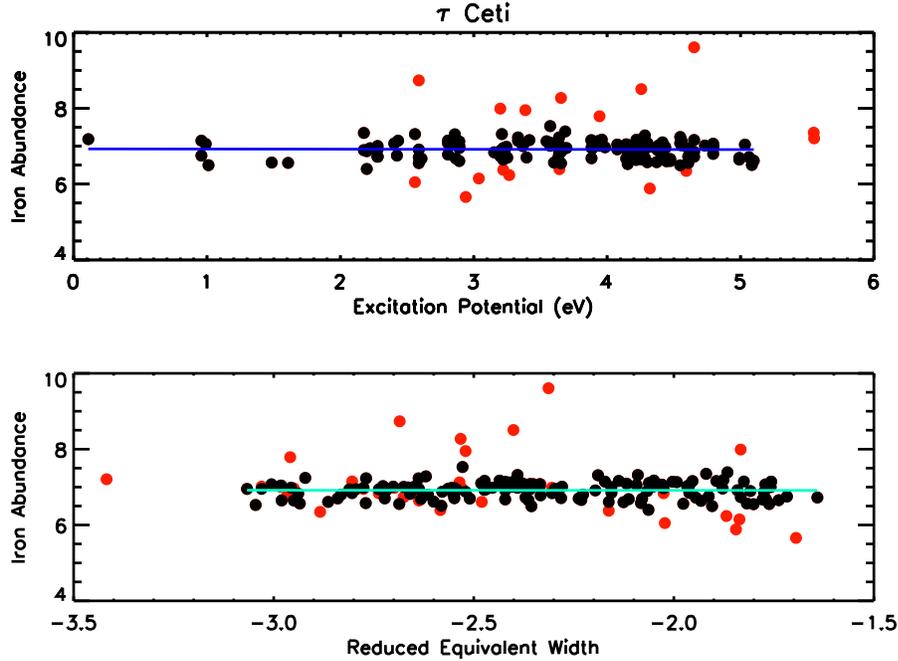}
%}
\caption[Plots of Iron abundance vs EP and Iron abundance vs Reduced EW for $\tau$~Ceti]{(Top panel) Iron abundance for $\tau$~Ceti is plotted against excitation potential for each Fe I or Fe II line from the line list. The blue line is the least-square fit having the minimum slope for the best-determined model temperature. (Bottom panel) Iron abundance is plotted against reduced EW and the blue-green line indicates the least-square fit having a minimum slope for best-determined value of $v_{\rm{micro}}$. The red points are the discarded points having standard deviation beyond 1 $\sigma$ (not considered for the fit). The best-fit determined parameters are: $T\rm{_{eff}}=5400$~K, ${\rm log}~g=4.4$, $v_{\rm{micro}}=0.2$~km s$^{-1}$ for $[Fe/H]=-0.5$}
\label{ew_temp}
\end{figure}

%------------------------------------------------------------------------------------------------
%  FIG 5
%------------------------------------------------------------------------------------------------

\begin{figure}[!ht]
\centering
\hbox{
\includegraphics[width=5.0in]{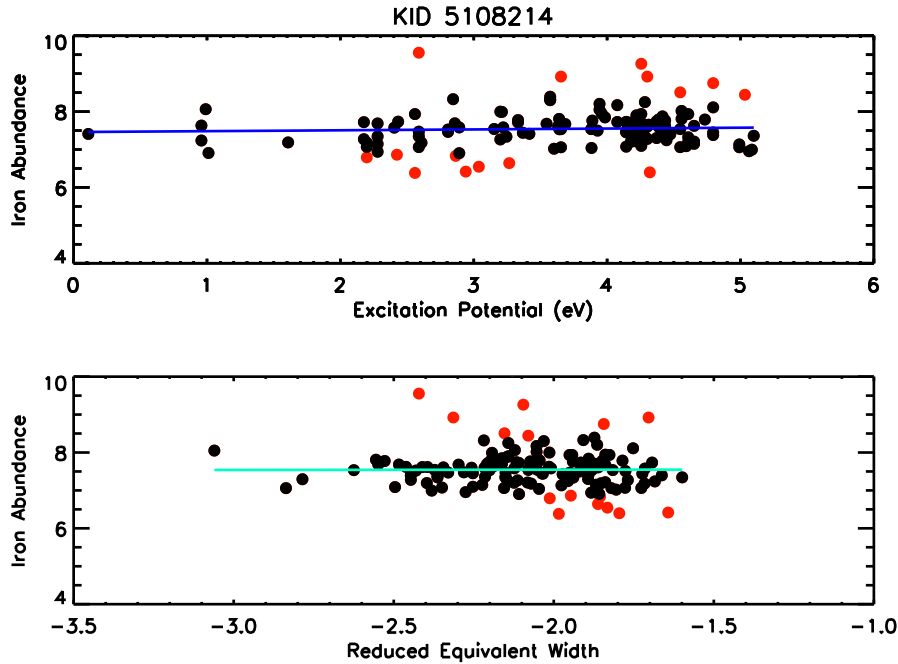}
}
\caption[A plot of Iron abundance vs EP and Iron abundance vs Reduced EW for KID~5108214]{(Top panel) Iron abundance for KID~5108214 is plotted against excitation potential for each Fe I or Fe II line from the line list. The blue line is the least-square fit having the minimum slope for the best-determined model temperature. (Bottom panel) Iron abundance is plotted against reduced EW and the blue-green line indicates the least-square fit having a minimum slope for best-determined value of $v_{\rm{micro}}$. The red points are the discarded points having standard deviation beyond 1 $\sigma$ (not considered for the fit). The best-fit determined parameters are: $T\rm{_{eff}}$=6050~K, $[Fe/H]$=0.5, ${\rm log}~g$ = $3.95$, $v_{\rm{micro}}$ = 1.65~km s$^{-1}$ for $[Fe/H]$ = 0.3}
\label{ew_temp_kid}
\end{figure}

%--------------------------------------------------------------------------------------------

The \textit{PARAS SPEC} routine is executed with a fixed value of metallicity. This value is taken from any previous measurements of metallicity done on the source as cited in literature or from the first method of synthetic spectral fitting. The abundances thus generated are plotted against EP and reduced EW for the respective lines. The ionization balance (Fe I - Fe II) is also estimated for each set of models. The slopes for the first two scatter plots and the difference of Fe I and Fe II are calculated for each set of models. The entire process is executed in two steps: first step on the coarse grid of models in $T\rm{_{eff}}$, ${\rm log}~g$ and $v_{\rm{micro}}$ and second step on the interpolated finer grid, similar to the previous method of synthetic spectral fitting. Thus, the model having a set of parameters where the slopes and the differences are simultaneously minimum gives us the best-determined $T\rm{_{eff}}$, ${\rm log}~g$ and $v_{\rm{micro}}$. 

As a test case, for star $\tau$~Ceti (SNR~$\sim$~500), the stellar parameters are determined by the EW method. The plot of iron abundance vs EP is shown in the upper panel of Fig.~\ref{ew_temp}. The figure also shows a solid line in blue obtained by a least-square fit having a slope of $0.0036_{-0.008}^{+0.007}$; indicative of the best-fit $T_{\rm{eff}}$ of 5400 K. In the bottom panel, a plot of iron abundance vs reduced EW is shown. A solid line in blue-green obtained by a least-square fit having a slope of $-0.002_{-0.001}^{+0.002}$ is shown indicative of best-fit $v_{\rm{micro}}$ of 0.2 km s$^{-1}$. The ${\rm log}~g$ value is determined where the difference between Fe I and Fe II abundances is minimum. Since the difference is just a number, no plot is shown for this case. For $\tau$~Ceti, the difference is $0.00_{-0.002}^{+0.002}$; indicative of a best-fit ${\rm log}~g$ value of 4.4. All the three parameters, $T_{\rm{eff}}$ and $v_{\rm{micro}}$ and ${\rm log}~g$ are obtained simultaneously where the slopes (Iron abundances vs EP and Iron abundances vs reduced EW) and difference of abundances of Fe~I and Fe~II are simultaneously minimum. A similar plot is shown for the star KID~5108214 having spectra of less SNR (SNR of $\sim$~100~pixel$^{-1}$) in Fig.~\ref{ew_temp_kid}. The upper panel of the figure also shows a solid line in blue obtained by a least-square fit having a slope of $0.018_{-0.006}^{+0.017}$ for iron abundances vs EP indicative of best-fit $T_{\rm{eff}}$ of 6050~K. In the bottom panel, a plot of iron abundance vs reduced EW is shown. A solid line in blue-green obtained by a least-square fit having a slope of $0.01_{-0.007}^{+0.004}$ indicative of best-fit $v_{\rm{micro}}$ of 1.65 km s$^{-1}$ is also shown in the bottom panel. Fe~I~-~Fe~II difference for KID~5108214 is given as $0.001_{-0.001}^{+0.009}$ indicative of a ${\rm log}~g$ value of 3.95.

%------------------------------------------------------------------------------------------------
%  FIG 6
%------------------------------------------------------------------------------------------------

\begin{figure}[!ht]
\centering
\vbox{
\includegraphics[height=4.0in]{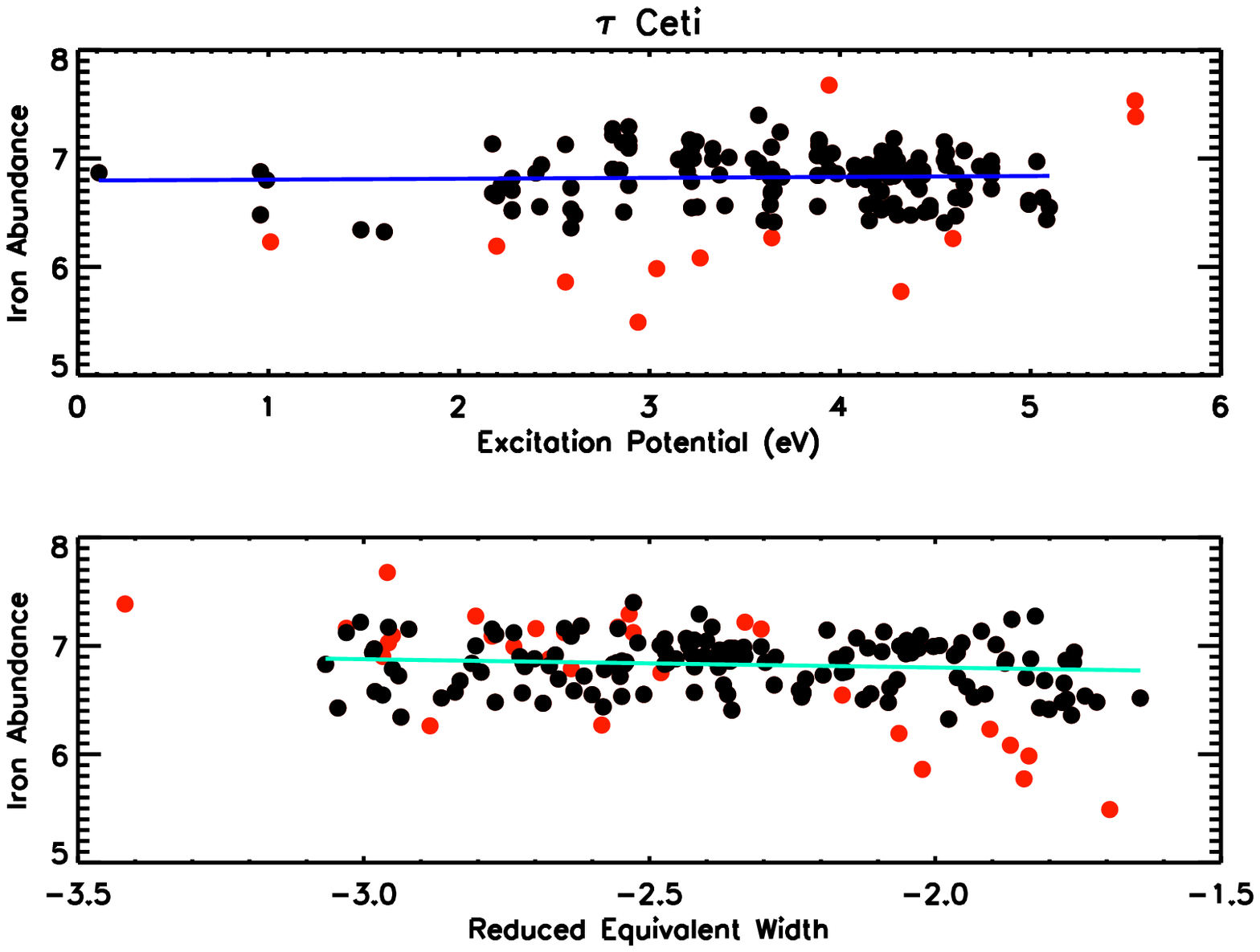}
%\vspace {1 in}
\includegraphics[height=4.0in]{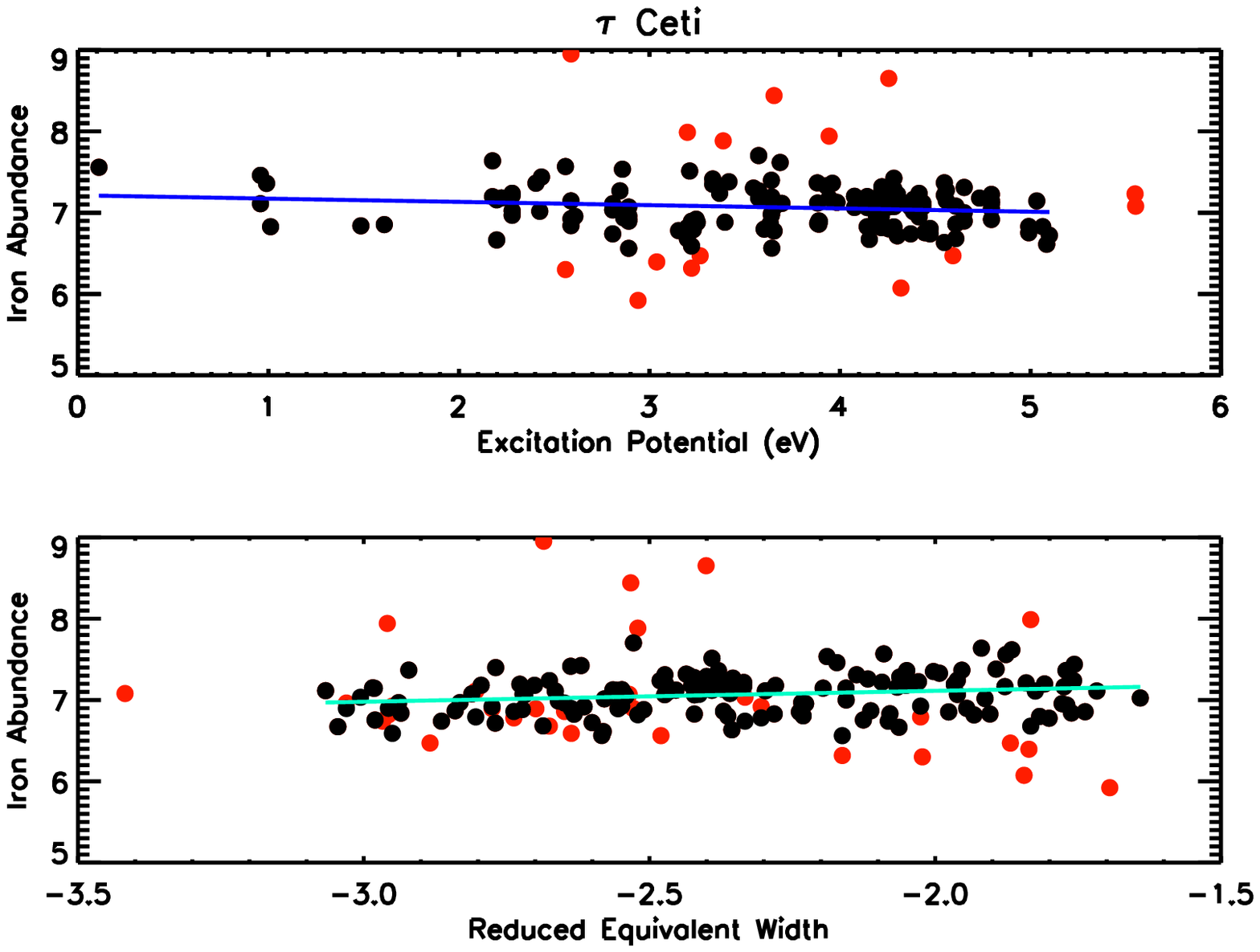}
}
\caption[Simulations done to demonstrate the effect of under-estimation and over-estimation of $T_{\rm{eff}}$ in \textit{PARAS SPEC}]{Simulations run for the star $\tau$ Ceti. The  $T\rm{_{eff}}$ for the upper two panels is 250 K lower than the correct $T_{\rm{eff}}$ and in the bottom two panels the $T_{\rm{eff}}$ is 250 K higher. Thus, we see a positive slope and negative slope in the abundance vs EP plot and Abundance vs Reduced EW in the upper two and lower two panels respectively.}
\label{testew}
\end{figure}

Both the parameters, $T\rm{_{eff}}$ and $v_{\rm{micro}}$, which are determined by fitting the slopes to the above-mentioned plot, are determined simultaneously. A slight positive or negative slope indicates under-estimation or over-estimation of $T_{\rm{eff}}$ and $v_{\rm{micro}}$ for the star respectively. Similarly, if the Fe I and Fe II difference is positive or negative, it indicates that ${\rm log}~g$ is under-estimated or over-estimated respectively. A set of simulations were done where we supplied models with higher $T\rm{_{eff}}$ or $v_{\rm{micro}}$ to see how the slope changes. The results of these simulations are shown in Fig~\ref{testew}. As a test case, we show here the simulations done in a step-size of 250~K for the value of $T\rm{_{eff}}$. This step-size chosen is just for illustrative purposes as it is the grid size of the original library with no interpolated models. This simulation is to give an idea how the other parameters of the star get affected if an under-estimated $T\rm{_{eff}}$ model is supplied. It is important to note that the synthetic spectra are finally interpolated on a finer grid to gain a better precision. When a lower $T\rm{_{eff}}$ model, 5150~K (250~K lower than the best-fit model) is supplied for the star $\tau$~Ceti, a positive slope ($0.008_{-0.01}^{+0.012}$ ) for Abundance vs EP plot is seen. The plot for abundance vs reduced EW has a slope of $-0.076_{-0.008}^{+0.006}$ and the difference of Fe~I and Fe~II is $-0.252_{-0.004}^{+0.004}$. Thus, we see that none of the parameters have achieved a minimum slope or a minimum difference indicating of a state where excitation equilibrium and ionization equilibrium is not achieved. For the case when a higher $T\rm{_{eff}}$ model, 5650~K (250~K higher than the best-fit model) is supplied, a negative slope for abundance vs EP plot is seen ($-0.04_{-0.009}^{+0.01}$). The plot for abundance vs reduced EW has a slope of $0.1346_{-0.006}^{+0.005}$ and the difference of Fe~I and Fe~II is $0.22_{-0.004}^{+0.003}$. The bottom panel, the slope of abundance vs Reduced EW also shows a departure from the zero slope. Similarly to the previous case, excitation and ionization equilibria are not achieved in this case too.

\section{Results}

We applied the \textit{PARAS SPEC} pipeline to the stars observed for the exoplanet and EB program run with PARAS. We also retrieved some of the F, G and K type stars observed for the \textit{Gaia} ESO survey from the SOPHIE archive~\footnote{http://atlas.obs-hp.fr/sophie/}. Both the methods of synthetic spectral fitting and EW method are applied on several known stars. We have compared the results with previous determination of stellar parameters. The results are summarized in 
Table~\ref{resultspectra}. The results for PARAS data are consistent within $1\sigma$ error bars of the literature values. For SOPHIE data, the estimated parameters agree within atmost $2\sigma$ of the quoted literature values. For one of the hot star HD~84937, the EW method was not applicable as for hot stars the spectral line blends make the EW estimation disfficult.

%------------------------------------------------------------------------------------------------

%------------------------------------------------------------------------------------------------
%  TABLE 2
%------------------------------------------------------------------------------------------------

\begin{landscape}
\begin{table}
\tabcolsep=0.15cm 
\caption{Results obtained from Spectral analysis (SF=Spectral Fitting; EW=Equivalent Width Method; LV= Literature Value (Errors represented on each parameter are model dependent as discussed in text). S/N is per pixel S/N at 6000~\AA. `P' stands for PARAS data and `S' stands for SOPHIE data.} 
\label{resultspectra}
\begin{tabular}{|l|c|c|c|c|c|c|c|c|c|c|c|c|r|}\\
\hline
\multicolumn{1}{|l|}{Star} &  \multicolumn{1}{|c|}{S/N} &   \multicolumn{1}{|c|}{Spectrograph} & \multicolumn{3}{|c|}{$T_{\rm{eff}}$} & \multicolumn{3}{|c|}{$[Fe/H]$} & \multicolumn{3}{|
c|}{${\rm log}~g$} & \multicolumn{1}{|r|}{$ v sin i $} & \multicolumn{1}{|r|}{Reference}\\
\hline
& & & SF & EW  & LV & SF & EW (fixed $[Fe/H]$) & LV & SF & EW  & LV & \\
\hline
Tauceti  & 500 & P & $5400 \pm 44$ & $5400 \pm 47$ & $5414 \pm 10$ & $-0.50 \pm 0.07$ & $-0.5$ & $-0.5 \pm 0.01$ & $4.40 \pm 0.09$ & $4.5 \pm 0.11$ & $4.49 \pm 0.03$ & $3.0 \pm 1.0$ & (a) \\
Sigma Draconis  & 250 & P & $5450 \pm 44$ & $5475 \pm 47$ & $5400 \pm 50$ &  $-0.1 \pm 0.07$ & $-0.1$  & $-0.20 \pm 0.06$ & $4.50 \pm 0.09$ & $4.50 \pm 0.11$ & $4.5 \pm 0.05$ & $3.0 \pm 1.0$ & (b) \\
Procyon & 550 & P & $6550 \pm 44$ & $6650 \pm 47$ & $6554 \pm 18$ &  $0.0 \pm 0.07$ & $0.0$  & $-0.04 \pm 0.01$ & $3.9 \pm 0.09$ & $3.9 \pm 0.11$ & $3.99 \pm 0.17$ & $5.0 \pm 1.0$ & (a) \\
HD9407 & 220 & P & $5700 \pm 44$ & $5725 \pm 51 $ & $5661 \pm30 $ & $0.0 \pm 0.11$ & $0.0$ & $0.03 \pm 0.09$ & $4.4 \pm 0.09$ & $4.35 \pm 0.11$ & $4.42 \pm 0.11$ &  $3.0 \pm 1.0$ & (c) \\
HD 166620  & 160 & P & $5200 \pm 62$ & $5025 \pm 51 $ & $4966 \pm 205 $ & $0.0 \pm 0.11$ & $0.0$ & $-0.17 \pm 0.08$ & $4.6 \pm 0.09$ & $4.35 \pm 0.14$ & $4.45 \pm 0.17$ & $4.0 \pm 1.0$ &  (c) \\
NLTT 25870  & 80 &P & $5400 \pm 84$ & $5225 \pm 67 $ & $5326 \pm 45 $ & $0.3 \pm 0.14$ & $0.3$ & $0.4 \pm 0.07$ & - & $4.6 \pm 0.14$ & $4.45 \pm 0.08$ & $3.0 \pm 1.0$ & (d) \\
HD 285507  & 60 & P & $4650 \pm 120$ & $4450 \pm 109 $ & $4542 \pm 50 $ & $0.1 \pm 0.14$ & $0.1$ & $0.13 \pm 0.05$ & - & -{$^{\star}$} & $4.67 \pm 0.14$ & $3.0 \pm 1.0$ &  (e) \\
KID 5108214 & 100 & P &$6000 \pm 84$ & $6050 \pm 67 $ & $5844 \pm 75 $ & $0.3 \pm 0.14$ & $0.3$ & $0.2 \pm 0.1$ & $4.0 \pm 0.14$ & $3.95 \pm 0.11$ & $3.80 \pm 0.01 $ & $5.0 \pm 1.0$ & (f) \\
HD 49674  & 90 & P &$5650 \pm 84$ & $5600 \pm 67 $ & $5632 \pm 31 $ & $0.2 \pm 0.11$ & $0.2$ & $0.33 \pm 0.01$ & - & $4.35 \pm 0.14$ & $4.48 \pm 0.12$ & $3.0 \pm 1.0$ & (g) \\
HD 55575 & 410 & P  &$5850 \pm 44$ & $5825 \pm 47 $ & $5850 \pm 70 $ & $-0.5 \pm 0.07$ & $-0.5$ & $-0.3 \pm 0.1$ & $4.30 \pm 0.09$ & $4.1 \pm 0.14$ & $4.2 \pm 0.15$ & $3.0 \pm 1.0$ & (b) \\
51~Ari & 120 & P  &$5700 \pm 62$ & $5700 \pm 51 $ & $5666 \pm 40 $ & $0.1 \pm 0.11$ & $0.1$ & $0.14 \pm 0.1$ & $4.30 \pm 0.09$ & $4.3 \pm 0.11$ & $4.45 \pm 0.1$ & $3.0 \pm 1.0$ & (b) \\

18~Sco & 150 & S & $5600 \pm 62$ & $5725 \pm 51$ & $5810 \pm 50$ & $0.0 \pm 0.11$ & $0.0$ & $0.01 \pm 0.05$ & $4.1 \pm 0.09$ & $4.25 \pm 0.11$ & $4.44 \pm 0.1$ & $3.0 \pm 1.0$ & (a) \\
61~Cyg~A & 215 & S & $4450 \pm 44$ & $4475 \pm 51$ & $4374 \pm 50$ & $-0.1 \pm 0.07$ & $-0.1$ & $-0.3 \pm 0.05$ & $4.3 \pm 0.09$ & $4.65 \pm 0.11$ & $4.63 \pm 0.1$ & $3.0 \pm 1.0$ & (a) \\
$\beta$~Gem & 577 & S & $5100 \pm 44$ & $4825 \pm 51$ & $4858 \pm 50$ & $0.3 \pm 0.07$ & $0.3$ & $0.12 \pm 0.05$ & $3.5 \pm 0.09$ & $2.75 \pm 0.11$ & $2.9 \pm 0.1$ & $4.0 \pm 1.0$ & (a) \\
$\beta$~Vir & 413 & S & $5900 \pm 44$ & $6050 \pm 51$ & $6083 \pm 50$ & $0.1 \pm 0.07$ & $0.1$ & $0.21 \pm 0.05$ & $3.6 \pm 0.09$ & $3.85 \pm 0.11$ & $4.1 \pm 0.1$ & $4.0 \pm 1.0$ & (a) \\
$\delta$~Eri & 583 & S & $5150 \pm 44$ & $4975 \pm 51$ & $4954 \pm 50$ & $0.1 \pm 0.07$ & $0.1$ & $0.06 \pm 0.05$ & $3.7 \pm 0.09$ & $3.55 \pm 0.11$ & $3.75 \pm 0.1$ & $3.0 \pm 1.0$ & (a) \\
Gmb~1830 & 284 & S & $4900 \pm 44$ & $5000 \pm 51$ & $4827 \pm 50$ & $-1.5 \pm 0.07$ & $-1.5$ & $-1.46 \pm 0.05$ & $4.8 \pm 0.09$ & $4.3 \pm 0.11$ & $4.6 \pm 0.1$ & $3.0 \pm 1.0$ & (a) \\
HD~22879 & 207 & S & $5650 \pm 44$ & $5700 \pm 51$ & $5868 \pm 50$ & $-0.8 \pm 0.07$ & $-0.8$ & $-0.88 \pm 0.05$ & $4.3 \pm 0.09$ & $4.0 \pm 0.11$ & $4.27 \pm 0.1$ & $3.0 \pm 1.0$ & (a) \\
$\mu$~Cas & 418 & S & $5550 \pm 44$ & $5475 \pm 51$ & $5308 \pm 50$ & $-0.6 \pm 0.07$ & $-0.6$ & $-0.82 \pm 0.05$ & $5.0 \pm 0.09$ & $4.5 \pm 0.11$ & $4.41 \pm 0.1$ & $3.0 \pm 1.0$ & (a) \\
{$^{\star \star}$}HD~84937 & 167 & S & $6400 \pm 62$ & -- & $6356 \pm 51$ & $-1.9 \pm 0.11$ & -- & $-2.09 \pm 0.05$ & $4.7 \pm 0.09$ & -- & $4.15 \pm 0.1$ & $10.0 \pm 1.0$ & (a) \\

\hline
\end{tabular}
%\vspace{0.5cm}
\scriptsize {References: (a) \cite{Blanco-Cuaresma2014}; (b) \cite{Soubiran2010}; (c) \cite{Paletou2015}; (d) \cite{Butler2000}; (e) \cite{McDonald2012}; \\
(f) KEPLER CFOP (https://cfop.ipac.caltech.edu); (g) \cite{Ghezzi2014}; (h) \cite{Cameron2007} \\}
{$^{\star}$} No Fe II lines were shortlisted for EW determination. \\
$^{\star \star}$ Hot star
\end{table} 
\label{resultspectra}
\end{landscape}

%---------------------------------------------------------------------------------------------------------------------------------------

\subsection{Error estimation and limitations of the method}

In the synthetic spectral fitting code, the parameters $T\rm{_{eff}}$, $[Fe/H]$ and ${\rm log}~g$ are fitted simultaneously. The parameter values are estimated by computing $\chi^2_{min}$ in a 3 parameter space. Errors on each of the parameters are computed by using constant $\chi^2$ boundaries as confidence limits on the three parameters jointly \citep{Press1992}. The errors shown in Table~\ref{resultspectra} are 68$\%$ confidence intervals for the parameters. The synthetic spectral fitting method yields reliable results only for spectra having S/N per pixel $\geqslant$~80. Between S/N $80-100$, the wavelength region of $6000-6500~\AA$~can be used for stellar property estimation. However, we lose information on ${\rm log}~g$ which is determined by Mg~I lines ($5160-5190~\AA$). For such cases, the EW method which works for spectra having S/N per pixel $\geqslant$~50 can be used to determine stellar properties. For, S/N $\geqslant$ 100, both the methods work well to determine all three stellar parameters, $T\rm{_{eff}}$, $[Fe/H]$ and ${\rm log}~g$. If the S/N per pixel at 6000~\AA~is above 120, the uncertainities in $T\rm{_{eff}}$ and ${\rm log}~g$ are $\pm25$ K and $\pm0.05$ respectively for both the methods. For S/N per pixel (at 6000~\AA~) between 80--100, the uncertainties are $\pm50$ K and $\pm0.1$ respectively. Similar numbers for S/N per pixel between 50--80 are $\pm100$ K and $\pm0.1$. Apart from these uncertainties, there could be systematic errors introduced for each of the methods. Detailed systematic error analysis for these methods is beyond the scope of the current work. Thus, we have referred to the similar work done by \cite{Blanco-Cuaresma2014}. As discussed in the paper, for synthetic spectral fitting method, there would be systematic errors due to different kind of models considered, different linelists used for the generation of synthetic spectra, and the choice of consideration of elements (iron or all elements) used for the fitting. All these factors on average give rise to uncertainties $\sim$37 K in $T\rm{_{eff}}$, $\sim$0.07 in ${\rm log}~g$ and $\sim$0.05 in $[Fe/H]$. We have further observed that there are additional systematic errors for stars having S/N per pixel less than 100 due to improper stellar continuum estimation which is of the order of the grid size of the synthetic spectra library. Moreover, there could be systematic errors introduced if a smaller wavelength region (6000--6500~\AA) (for S/N $<$ 100) is used for the estimation of stellar parameters instead of the entire wavelength region (5050--6500~\AA). Such uncertainties are $\sim$50 K in $T\rm{_{eff}}$ and $\sim$0.1 dex in ${\rm log}~g$. Considering all these factors, the systematic uncertainties for stars having S/N less than 100 is around 67 K in $T\rm{_{eff}}$, 0.11 in ${\rm log}~g$ and 0.11 in $[Fe/H]$ in case of spectral fitting method. 

As mentioned earlier, EW of a line is determined by modelling the spectral profile of the absorption feature of the star by MPFIT function in {\tt{IDL}} \citep{Markwardt2009} as shown in Eq.~\ref{eq2}. The absorption lines are modelled by Gaussian profile and 1~$\sigma$ errors on each of the fitting parameters of the function are obtained as given in \cite{Markwardt2009}. These errors are used to compute the 1~$\sigma$ error bars on the obtained EW values for each of the absorption lines. Elemental abundances from the lines are derived using EW values and a stellar model (function of $T\rm{_{eff}}$, $[Fe/H]$ and ${\rm log}~g$) as inputs as shown below. We determine three set of abundances, one on the original set of EWs, one on EW+$\sigma_{EW}$ and the third set on EW-$\sigma_{EW}$. The sets of abundances determined on EW+$\sigma_{EW}$ and EW-$\sigma_{EW}$ correspond to the two extreme values of abundances determined as shown in Eq.~\ref{eq3a} and \ref{eq3b}.

\begin{equation}\label{eq3a}
\centering
Abundance_{+\sigma} = Abundance (EW + \sigma_{EW})
\end{equation}

\begin{equation} \label{eq3b}
\centering
Abundance_{-\sigma} = Abundance (EW - \sigma_{EW}) 
\end{equation}

For each set of Abundances, Abundance$_{+\sigma}$ and Abundance$_{-\sigma}$, and different stellar models we derive a set of best-fit parameters, $T\rm{_{eff}}$, $[Fe/H]$ and ${\rm log}~g$ in the same way as described earlier. This provides an upper and lower-limits on each of the best-fit parameters of $T\rm{_{eff}}$, $[Fe/H]$ and ${\rm log}~g$.
EW method could also have systematic errors due to different kinds of models and linelists used for the estimation of EWs, and due to the rejection of outliers during linear fitting for the slope determination. These systematic errors as interpreted from \citep{Blanco-Cuaresma2014} are of the order of 45 K in $T\rm{_{eff}}$ and 0.1 in ${\rm log}~g$. The stellar parameters derived along with their combined formal and systematic uncertainties for each of the methods are listed in Table~\ref{resultspectra}.
Fast rotating stars will have line blends and will pose difficulty in measuring accurate EW of the lines. Visual inspection for all the lines, despite being a cumbersome task, is necessary to cross-check and discard line blends and improper fits. Thus, this method can be applied on stars having low SNR/pixel between ($50-80$) unlike the synthetic spectral fitting method for determination of stellar parameters though with larger error bars. 

\cite{Blanco-Cuaresma2014} applied the synthetic spectral fitting and EW methods on many stars. It is important to note that these are average errors reported on an average SNR/pixel of 50. Thus, errors given in Table~\ref{resultspectra} only correspond to the fitting errors and are formal errors on each parameter. As quoted by \cite{Smalley2005}, realistically the typical errors on the atmospheric parameters of a star determined by any method are of the order of $\pm$100~K for $T_{\rm{eff}}$, $\pm$0.1~dex for $[Fe/H]$ and $\pm$0.2~dex for ${\rm log}~g$. The exact magnitude of the uncertainty will depend upon the sensitivity of the lines used in the analysis. \cite{Mortier2014} have compared the accuracies of the atmospheric parameters obtained by photometry, spectroscopy and asteroseismology and find asteroseismology to yield the most reliable results. 

\section{Conclusions and Future Scope}\label{concl}

The stellar parameters derived from the code \textit{PARAS SPEC} agree well with those reported in literature as indicated in Table~\ref{resultspectra}. \cite{Torres2010} studied a large sample of EBs which have masses, radii and stellar parameters studied at a precision as high as 3$\%$. From this study, the authors reported an empirical relationship of mass and radius of stars above 0.6~M$_{\odot}$ as a function of the  $T\rm{_{eff}}$, $[Fe/H]$ and ${\rm log}~g$ as follows:
\begin{equation} \label{3.5}
\centering
logM=a_1+a_2X+a_3X^2+a_4X^3+a_5(log~g)^{2}+a_6(log~g)^{3}+a_7[Fe/H]
\end{equation}

\begin{equation} \label{3.6}
\centering
logR=b_1+b_2X+b_3X^2+b_4X^3+b_5(log~g)^{2}+b_6(log~g)^{3}+b_7[Fe/H]
\end{equation}

where X~=~log$(T\rm{_{eff}})-4.1$. The calibration coefficients a$_i$ and b$_i$ are given in \cite{Torres2010}. The reliability of stellar parameters derived from high-resolution spectroscopy is expected to be better than compared by modelling photometry data \citep{Casagrande2011}. Thereby, this code can be used to deduce the masses and radii of primary stars at higher accuracies leading to precise determination of stellar parameters of the secondary component. 

We have successfully demonstrated the \textit{PARAS SPEC} code which is automated, easy to use and can handle data from different high resolution spectrographs. It also has the provision of working with low S/N data by stacking multiple epochs to enhance the S/N of the combined spectra. This code is currently suitable to derive atmospheric parameters of F, G and K type stars. The Kurucz models used here are not suitable for giant stars. Thus, there is a furure plan to adapt different stellar atmosphere models like MARCS models~\footnote{http://marcs.astro.uu.se/} and BT-Settl models \citep{Baraffe1998}. There are also plans to develop this code for low mass stars by working with the NIR linelist of {\tt{SPECTRUM}}. BT-Settl models suit this requirement. \textit{PARAS SPEC} will be updated and expanded on these considerations.

\section*{Acknowledgments}

This work has been made possible by the PRL research grant for PC (author) and the PRL-DOS (Department of Space, Government of India) grant for PARAS. We acknowledge the help from Vaibhav Dixit, Vishal Shah and Aashka Patel for their technical support during the course of data acquisition. We also acknowledge help from Suvrath Mahadevan and Arpita Roy for the reduction and data analysis pipeline for PARAS. This research has made use of the ADS and CDS databases, operated at the CDS, Strasbourg, France and SOPHIE archival database team for the reduced SOPHIE data.

%\bibliographystyle{mnras}
%\bibliography{reference} % if your bibtex file is called example.bib

\end{document}